\documentclass[conference]{IEEEtran}
 \IEEEoverridecommandlockouts
\usepackage{cite}
\usepackage{amsmath,amssymb,amsfonts}
\usepackage{algorithmic}
\usepackage{graphicx}
\usepackage{textcomp}
\usepackage{comment}
\usepackage{adjustbox}
\usepackage{xcolor}
\usepackage{booktabs}
\usepackage{soul}
\def\BibTeX{{\rm B\kern-.05em{\sc i\kern-.025em b}\kern-.08em
    T\kern-.1667em\lower.7ex\hbox{E}\kern-.125emX}}

\begin{document}

\title{A Wide Dynamic Range Read-out System For Resistive Switching Technology\\

}

\author{
    \IEEEauthorblockN{Lijie Xie$^{*}$, Jiawei Shen$^{*}$, Andrea Mifsud$^{*\dagger}$, Chaohan Wang$^{*}$, Abdulaziz Alshaya$^{*}$, Christos Papavassiliou$^{*}$}
    \IEEEauthorblockA{$^{*}$Department of Electrical and Electronic Engineering, Imperial College London, SW7 2BT, UK\\$^\dagger$Centre for Bio-Inspired Technology, Institute of Biomedical Engineering, Imperial College London, SW7 2AZ, UK\\
    Corresponding author email: lijie.xie19@imperial.ac.uk
    }%
}%

\maketitle

\begin{abstract}



The memristor, because of its controllability over a wide dynamic range of resistance, has emerged as a promising device for data storage and analog computation. A major challenge is the accurate measurement of memristance over a wide dynamic range. In this paper, a novel read-out circuit with feedback adjustment is proposed to measure and digitise  input current in the range between 20nA and 2mA. The magnitude of the input currents is estimated by a 5-stage  logarithmic current-to-voltage amplifier which scales a linear analog-to-digital converter. This way the least significant bit tracks the absolute input magnitude. This circuit is applicable to reading single memristor conductance, and is also preferable in analog computing where read-out accuracy is particularly critical. The circuits have been realized in Bipolar-CMOS-DMOS (BCD) Gen2 technology. 

\end{abstract}


\section{Introduction}

Memristors, known as resistive memory devices, were conceptually proposed by Chua in 1971\cite{chua}.
In theory, the resistance of the two-terminal memristor can be adjusted by voltages applied on it. 
This theory came into reality in 2008 with the birth of the first memristor\cite{HP}, which works as an access device in a crossbar. Since then, memristors have drawn substantial interest, developing from atomic switches\cite{atomic_switch} to random resistive access memory (RRAM)\cite{RRAM} and multi-bit memory storage\cite{multi_bit}. Besides, on-going tests on employing memristors as electronic synapses/processors in neural networks\cite{face}\cite{fully2}\cite{image} manifest the capability of in-memory computing. 


However, challenging issues are still encountered in precisely reading-out the memristor conductance or the current flowing through memristors. Firstly, only a low read voltage can be applied to the memristor to ensure minimal changes to its conductance. Although \cite{multi_bit} allows a large read voltage with compensation pulses for the conductance restoration, this approach works only for low-bit precision. Thus most read-outs in analog computing still require a low read voltage. 
Secondly, memristors have a large current range of multiple orders of magnitude, which is a great challenge for conventional read-out circuits. \cite{efficient} utilises a transimpedance amplifier (TIA) to provide a virtual ground and realizes I-to-V conversion. However, if the input current is extremely low then the converted voltage is too small to be read with high accuracy; If the current is extremely high, then the TIA cannot guarantee that the virtual ground is stable. Besides, the requirement for reading accuracy is made stricter in neural networks where memristors are performing dot-product calculation\cite{dot-product}. The dot-product in neuromorphic networks implements analog computing, obtaining the exact current value as the result. Recent approaches\cite{fully}\cite{hardware} employ the integrator for I-to-V conversion, but the integrated voltage inevitably reduces the read voltage and jeopardises computing performance. In consequence, a dynamic read-out circuit which can be adaptable to various conductance orders of magnitude is highly desired in memristor applications.

\begin{figure}[!t]
  \centering
  \includegraphics[scale=0.175]{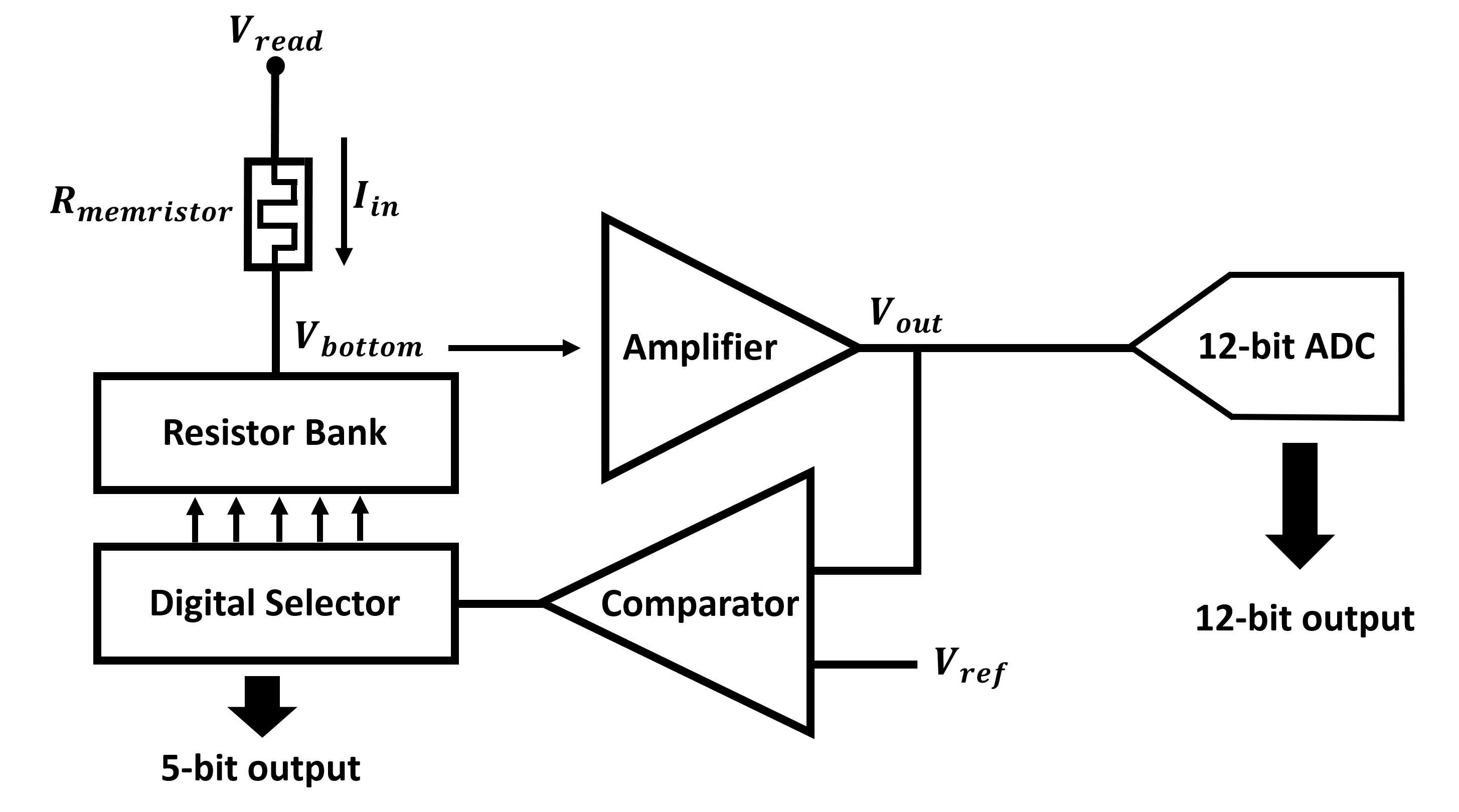}
  \caption{\footnotesize Read-out block diagram. The input current is read out as the combination of a 5-bit output estimating the input magnitude, and a 12-bit absolute value. The magnitude estimation is achieved by a feedback operation. }
  \label{Fig:1}
\end{figure}

This paper provides a novel method to read out and digitise the input current flowing through the measured memristor with high accuracy and linearity under feedback control. An input current ranging from 20nA to 2mA can be converted to a voltage bounded in an appropriate range. This is achieved by using the logarithmic resistor bank which has 5 logarithmic resistance values corresponding to 5 orders of input current magnitudes. The feedback control can select the most suitable resistor for I-to-V conversion. The converted voltage is amplified to meet the input range of the analog-to-digital converter (ADC) and then be digitised. This read-out method performs an adaptable I-to-V conversion and offers negligible corruption to the read voltage between the measured memristor, minimising error introduced by the measurement circuit. This read-out circuit is a promising candidate in both measuring single memristor conductance and processing of dot-product engine. Section~\ref{1} provides the architecture of the read-out circuit and the principle of design implementation. Section~\ref{2} presents the simulation result analysis and the circuit layout.

\section{Design Overview}
\label{1}


The monolithic circuit (Fig. \ref{Fig:1}) consists of a memristor to be measured, a resistor bank, a switched capacitor amplifier, a comparator, a digital selector, and an ADC. When the read-out circuit operates, a read voltage $V_{read}$ is applied at the top electrode of the measured memristor. The bottom electrode of the memristor is connected to a resistor bank which has 5 logarithmic resistors. The selected resistor is negligible compared to the memristor resistance, minimising disruption to the input current
$I_{in}=\frac{V_{read}-V_{bottom}}{R_{mem}}$.
This resistor bank, with the lowest resistor selected initially, converts the input current to a voltage
$V_{bottom}$=$I_{in}\times{R_{bank}}$ 
which is then amplified by the switched capacitor amplifier. The amplified voltage $V_{out}$ needs to be compared with a reference voltage via the comparator, checking if $V_{out}$ reaches the input range of ADC. If $V_{out}$ is higher than the reference voltage, then $V_{out}$ and the selected resistor can be digitised. Otherwise, the digital selector is triggered to activate the selection of the higher resistor. 
By tuning up the resistance of the resistor bank, a higher $V_{out}$ can be obtained and sent to be compared. This process repeats until $V_{out}$ is high enough to be digitised. 

\begin{figure}[!b]
  \centering
  \includegraphics[scale=0.285]{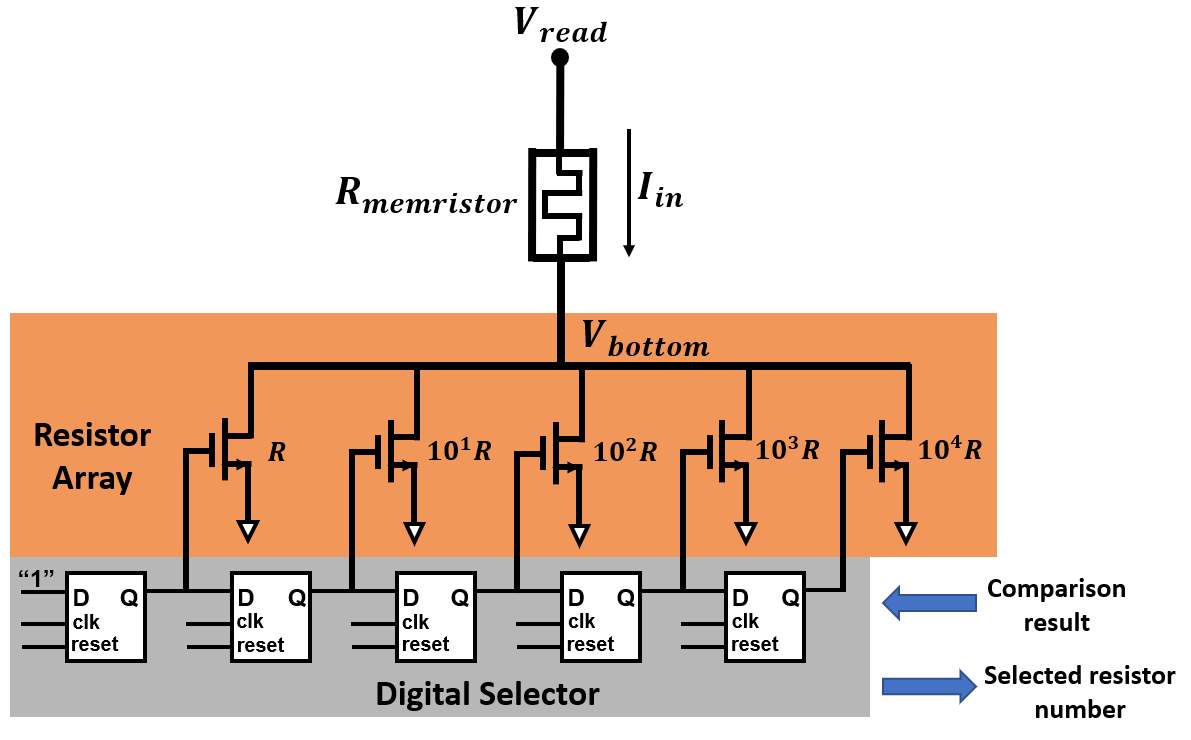}
  \caption{\footnotesize Schematic of the resistor bank (orange) which contains 5 NMOS resistors. Only 1 resistor can be selected at a time and the default one is R. The comparison result determines if the selected resistor matches the input magnitude, or a larger resistor should be switched on to offer larger I-to-V gains.
 }
  \label{Fig:2}
\end{figure}

\subsection{Resistor Bank}

The resistor bank (Fig. \ref{Fig:2}) contains 5 NMOS resistors working in triode region. These resistors offer a dynamic I-to-V conversion.
NMOS resistors are preferred than passive resistors due to the following reasons. Firstly, the NMOS transistor does not need extra switches as it intrinsically is a switchable resistor. Thus it can be reliably switched between off-resistance (high) and on-resistance (low) which can be controlled as $R_{on}=\mu_{n}C_{ox}(W/L)$
$(V_{GS}-V_{TH})^{-1}$. 
Secondly, extra switches are detrimental to the read-out accuracy because the read voltage inevitably decreases due to the voltage drop across the switch. The decrease is significant when the input current is so large that much of the read voltage is applied to this switch.
Besides, NMOS resistors occupy less chip area compared to their passive counterpart. Although NMOS transistors are easier affected by foundry process that the practical on-resistance may suffer from fabrication mismatch, NMOS transistors are still suitable resistors in this design. This design does not strictly require the resistance to be absolutely accurate because resistors are used to define dynamic measuring scales. In addition, the deviated on-resistance can be compensated back by varying $V_{GS}$ and other calibrations.

By fine-tuning the transistor size, five logarithmic resistance states around R, 10R, $10^2$R, $10^3$R, $10^4$R are required. Only one resistor can be selected at a time and by default the lowest resistor is selected first. Starting from the lowest value, the bottom voltage of the memristor is sent to be amplified and compared. If the amplified voltage is less than the ADC input threshold (157.3mV), the negative comparison result enables the digital selector to work. The digital selector, formed by shift registers, switches off R and switches on 10R. If the comparison result is still negative, $10^2$R is selected. This process continues until the positive result is given. The positive result terminates the resistor switching process and outputs the 5-bit resistor number.


\subsection{Switched Capacitor Amplifier}
\begin{figure}[!t]
  \centering
  \includegraphics[scale=0.2]{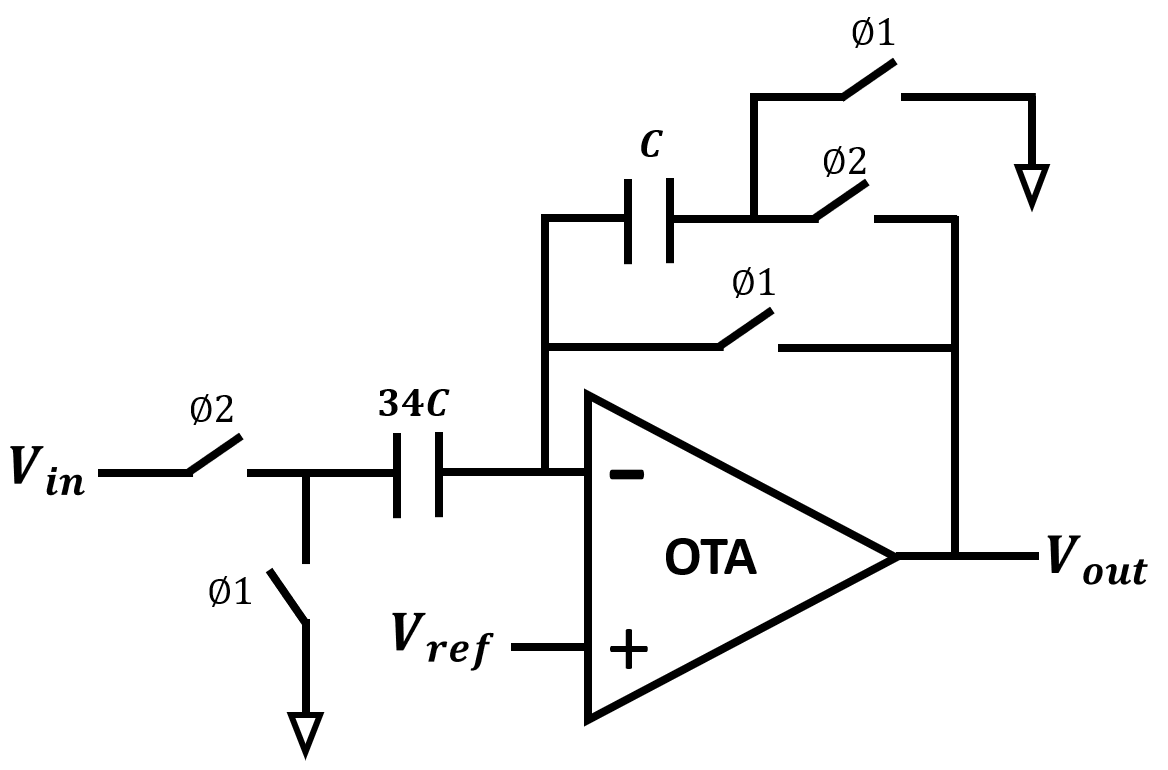}
  \caption{\footnotesize Schematic of the amplifier. The gain is obtained by the capacitor ratio.  }
  \label{Fig:3}
\end{figure}
\begin{figure}[!b]
  \centering
  \includegraphics[scale=0.14]{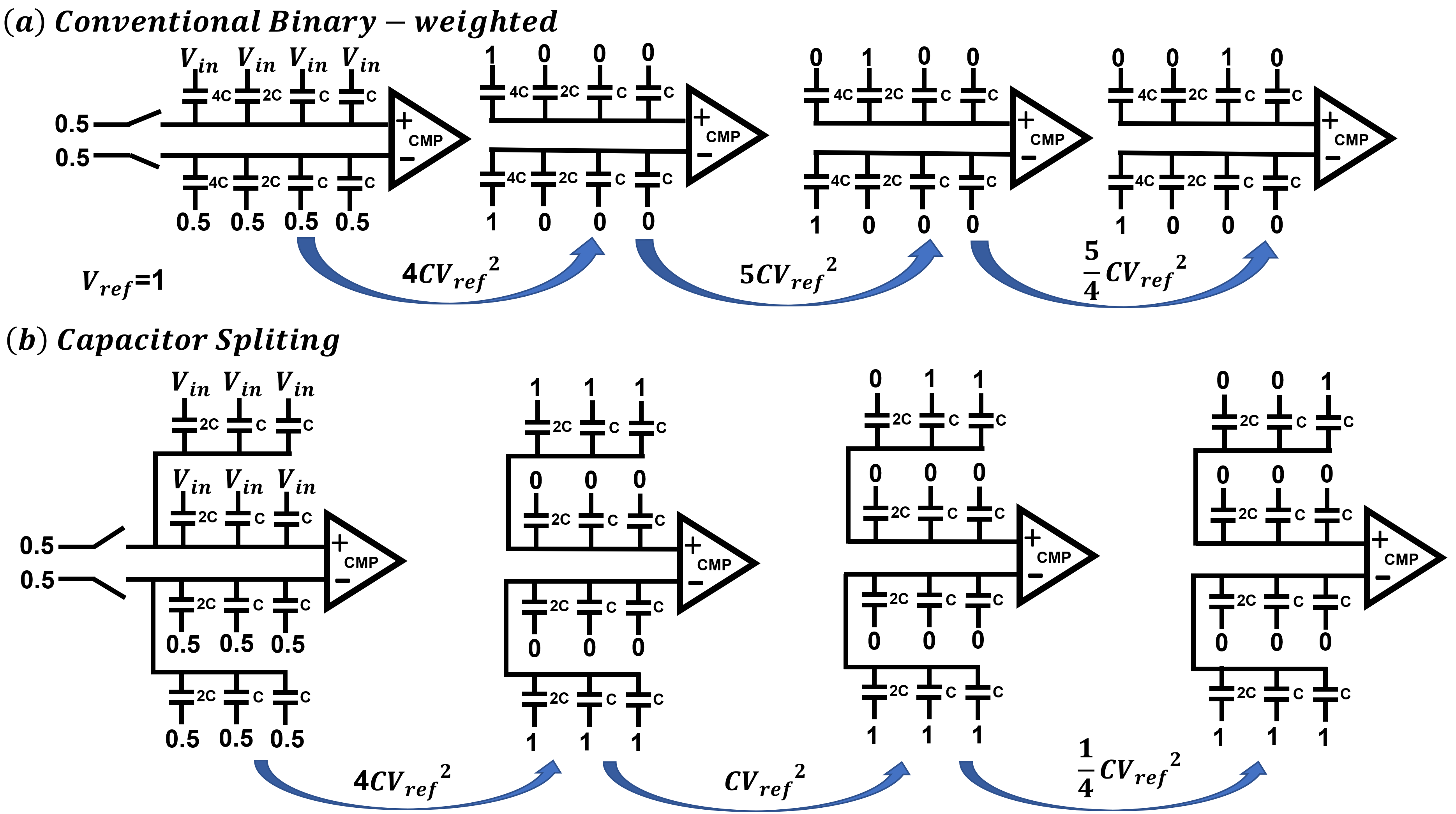}
  \caption{\footnotesize A simplified 4-bit comparison on structure, switching sequence, and power consumption in "down" transition. (a) Conventional binary-weighted structure. (b) Split "MSB" structure }
  \label{Fig:4}
\end{figure}

For precisely amplifying the low voltage $V_{bottom}$ to a relatively large value thus utilising the full input range of the ADC, a switched capacitor amplifier (Fig. \ref{Fig:3}) is exploited. 
A gain around 34 is offered by the capacitor ratio which can be accurately controlled with appropriate sizing and layout methodology. Two-phase switches operate to suppress the charge accumulated at the input node of the OTA, minimizing the amplification error.
Besides, this amplifier can be directly connected to the resistor bank with negligible interference due to the insulation of the capacitor.


\subsection{12-bit Analog-to-digital Converter}

A 12-bit SAR ADC is designed to digitise the input voltage ranging from 0.1V to 1.7V. This proposed ADC employs a differential architecture and capacitor array bottom-plate sampling to offset and reduce parasitic capacitance. For suppressing the charge/discharge power consumption as well as kickback and thermal noise, a split capacitor array structure with a unit capacitor of $\approx$30fF is utilised.

The energy efficiency, conversion speed, and digitisation precision can be substantially improved on each down (discharge) transition. This power saving (37$\%$) is achieved by splitting the most significant bit capacitor into the copy of remaining capacitors\cite{ADC} with the cost of more control signals needed. The simplified 3-bit comparison between this proposed method and the conventional counterpart is shown in Fig. \ref{Fig:4}. Take the first conversion as an example, the conventional C2C structure needs to discharge the 4C capacitor (4C$V_{ref}^{2}$) and then charges the 2C capacitor (C$V_{ref}^{2}$), with total power consumption $E_{1}$=5C$V_{ref}^{2}$; However, the proposed structure only needs to discharge a 2C capacitor ($E_{2}$=C$V_{ref}^{2}$) to finish the charge redistribution.

\section{Results}
\label{2}

\subsection{Memristance-to-voltage Conversion}\label{AA}
\begin{figure}[!b]
  \centering
  \includegraphics[scale=0.19]{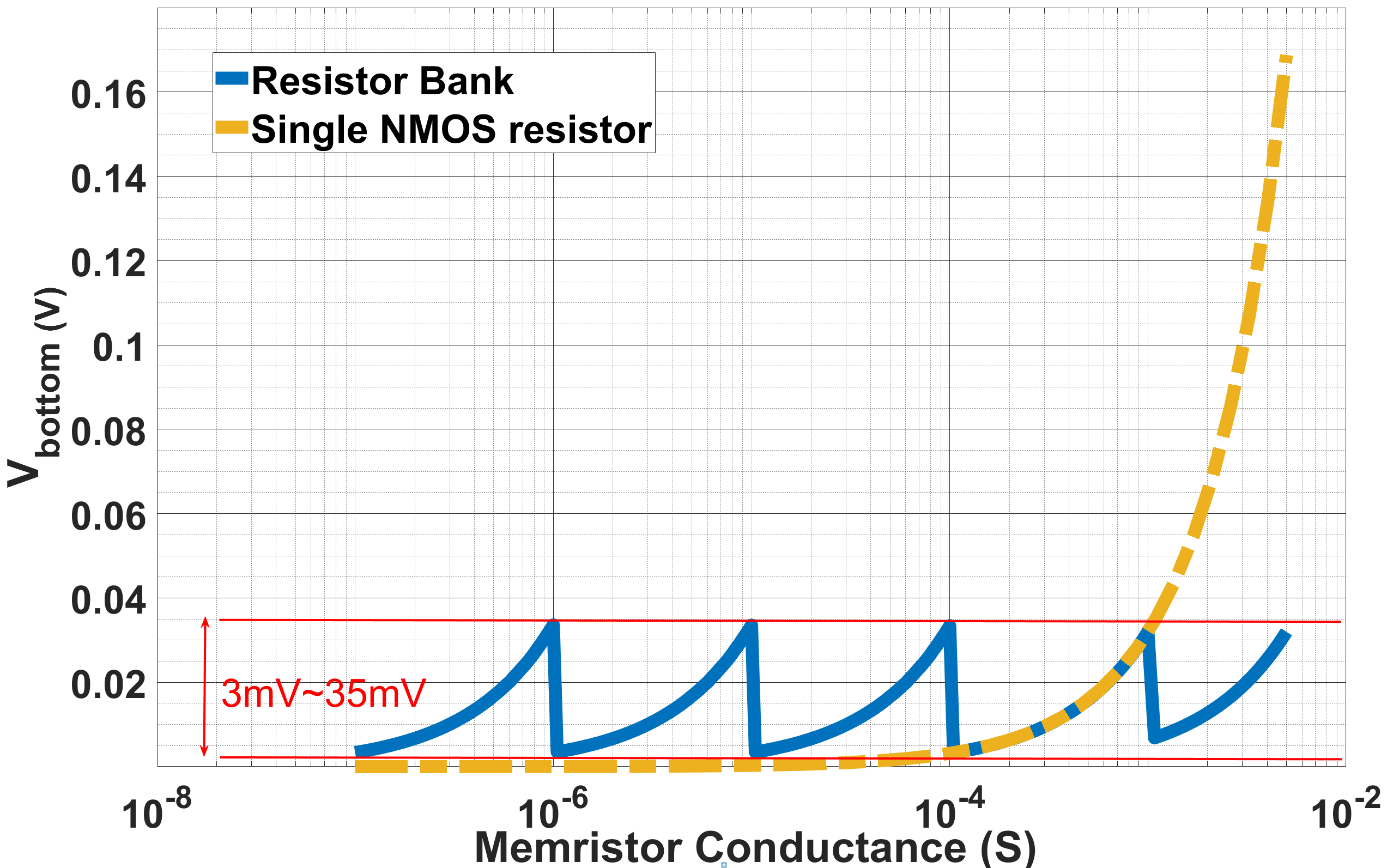}
  \caption{\footnotesize Memristance-to-voltage conversion. This work (blue): converted results are between 3mV and 35mV; Conventional design (yellow): extremely low/high converted voltage when input is low/high. 
  }
  \label{Fig:5}
\end{figure}
This section gives the improved current-to-voltage conversion performance.
The traditional one-resistor method for I-to-V conversion uses a large resistor if the input current is too small. However, this large resistor inevitably affects the conversion accuracy when the input current is large. To address this issue, the resistor bank is used, showing improved accuracy and linearity explained below.

The improved accuracy is due to the bounded and relatively low voltage level at the bottom terminal of the memristor. It avoids that the memristor loses much read voltage when the resistor is large. The comparison between employing this proposed resistor bank and using a single 10R NMOS resistor is shown in Fig. \ref{Fig:5}. With a 0.2V read voltage applied at the top terminal of the measured memristor which varies substantially from 100nS to 5mS, the bottom plate of the memristor is bounded between approximately 3mV and 35mV.This bounded voltage level minimises the variation of voltage drop between the memristor, and reduces the impact of $V_{ds}$ variation on NMOS resistance.

\begin{figure}[!t]
  \centering
  \includegraphics[scale=0.19]{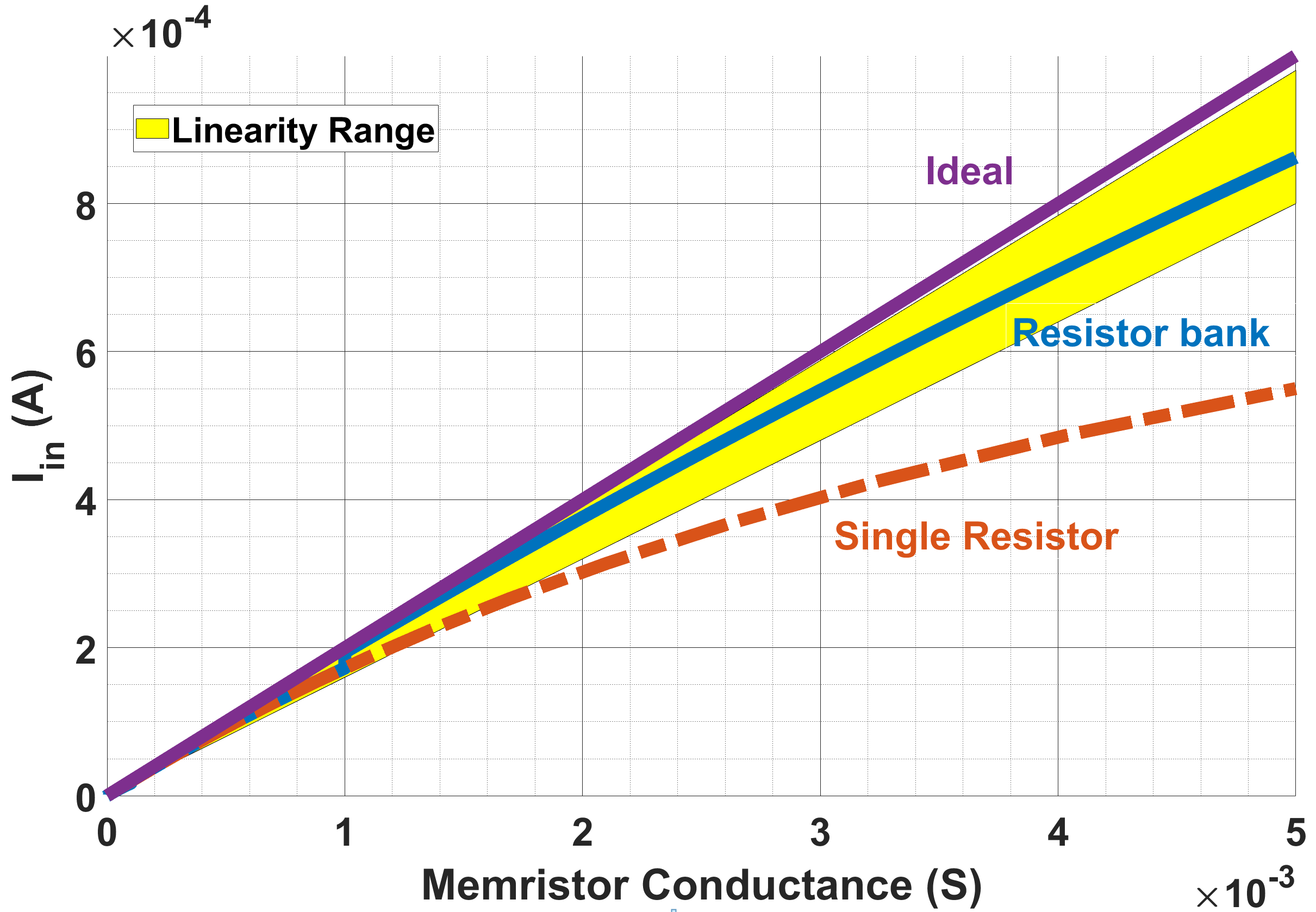}
  \caption{\footnotesize Memristance-to-current linearity. Ideally (purple), the converted current is perfectly linear with memristor conductance; This design (blue) drops only a little linearity, and the linearity range is within yellow; The conventional design (orange) greatly loses linearity.
  }
  \label{Fig:6}
 \end{figure} 
  
  \begin{figure}[!b]
  \centering
  \includegraphics[scale=0.4]{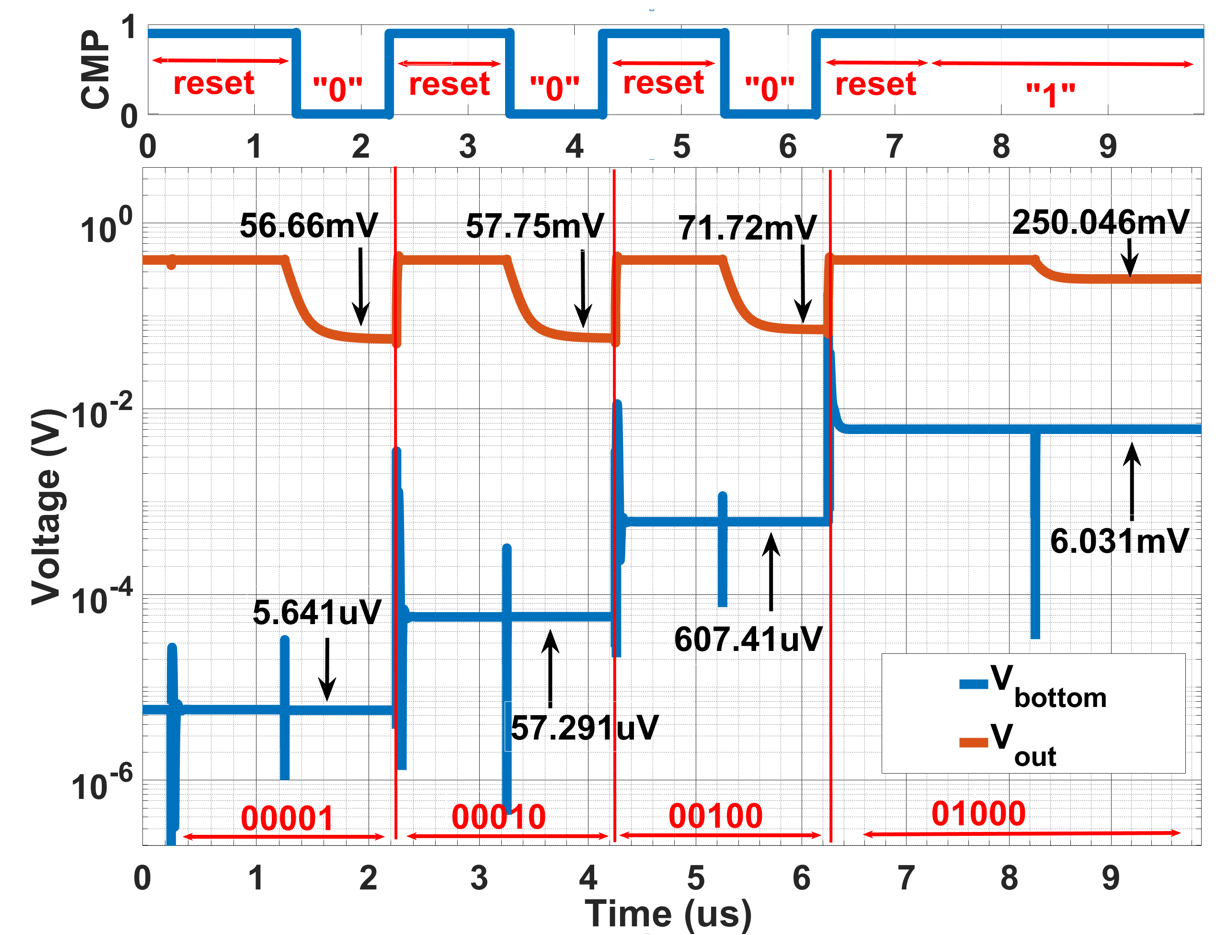}
  \caption{\footnotesize
  Single I-to-V timing. A randomly selected input current (355.66nA) is successively converted ($V_{bottom}$) and amplified ($V_{out}$, approximately $V_{out}$=$V_{cm}+A\times V_{bottom}$, $V_{cm}$ is around 56.54mV). With the comparison result (CMP), the second largest resistor (01000) is determined.
  }
  \label{Fig:7}
\end{figure}

The linearity of memristance-to-current conversion (Fig. \ref{Fig:6}) is also improved significantly by utilising the resistor bank. Inevitably the practical linearity (blue) is worse than the ideal one (purple) as the voltage drop across the memristor is not exactly $V_{read}$ and it varies with the input. Yellow gives a linearity range of the proposed design, the boundary (upper/lower) is met when $V_{read}$ is 3mV/35mV. The resistor bank can still maintain a high linearity compared to the single NMOS resistor (orange) which continuously loses linearity with growing memristor conductance.

\subsection{Successive Resistor Control Timing}

This section gives the operation of one single read-out process.
Fig. \ref{Fig:7} presents the successive resistor control in the resistor bank. An example of 355.66nA input is given. In this I-to-V process, $V_{bottom}$ starts from the lowest value (5.641µV) with the lowest resistance (00001) selected by default. As a 10 times larger resistor (00010) is chosen in next stage, the converted voltage (57.291µV) increases almost 10 times. The amplified voltage also increases as $ \Delta V_{out} = A \times \Delta V_{bottom}$.
In this case, the process continues until the second largest resistor (01000) is controlled. At this moment, the converted voltage (6.031mV) is amplified to 250mV, which is above the threshold voltage 157.3mV and ready for digitisation.
Besides, the successive searching time varies from 1 to 5 cycles. Fewer cycles are needed for deciding the resistor if the input current is larger.

\subsection{Full-range Digitisation}

\begin{figure}[!t]
  \centering
  \includegraphics[scale=0.19]{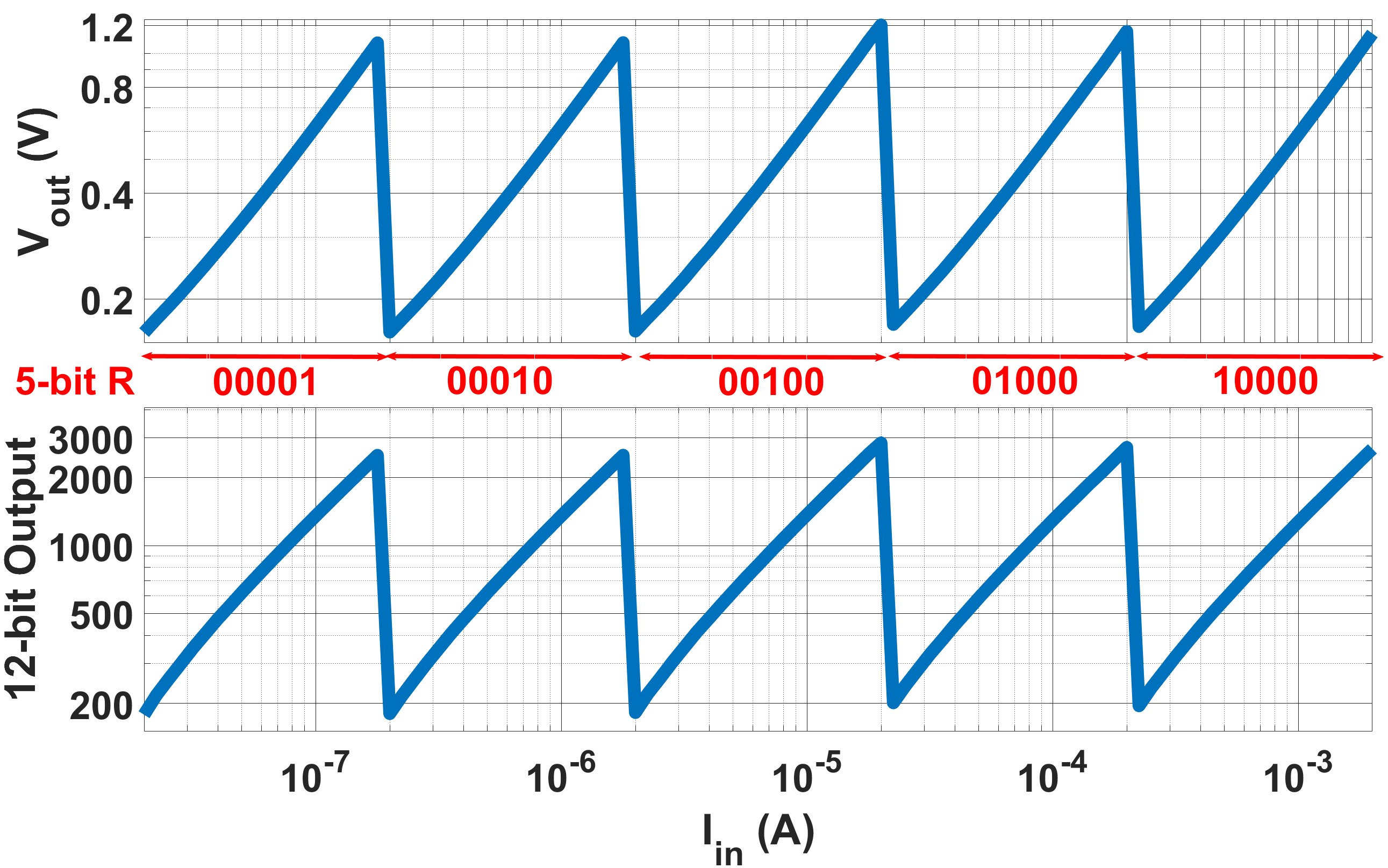}
  \caption{\footnotesize Full-range conversion/digitisation. Input currents between 20nA and 2mA can be converted and amplified to voltages between 160mV and 1.2V (upper), with different resistors (red) controlled for I-to-V conversion. Lower is the digitised result.
   }
  \label{Fig:8}
\end{figure}

This section provides the final result for full-range input currents.
Upper of Fig. \ref{Fig:8} shows that the amplified output voltage varies between 160.5mV and 1.205V as the input current ranges from 20nA to 2mA. This current range is logarithmically divided into 5 domains with each domain corresponding to a different resistor. 
Lower of Fig. \ref{Fig:8} provides the digitised result which can be used as a look-up table for measuring the input current. Each 12-bit digital output, assisted with a 5-bit resistor representation, reflects one exact input current. Every domain performs similar output ranges, monotonicity and linearity, guaranteeing that the input current can be digitised precisely.


$V_{out}$ variability due to process and local mismatch is around 20$\%$. So the input of ADC is set from 0.1V to 1.7V to give extra input margin. In addition, this error can be eliminated in post-calibration.



\subsection{Layout}
\begin{figure}[!t]
  \centering
  \includegraphics[scale=0.135]{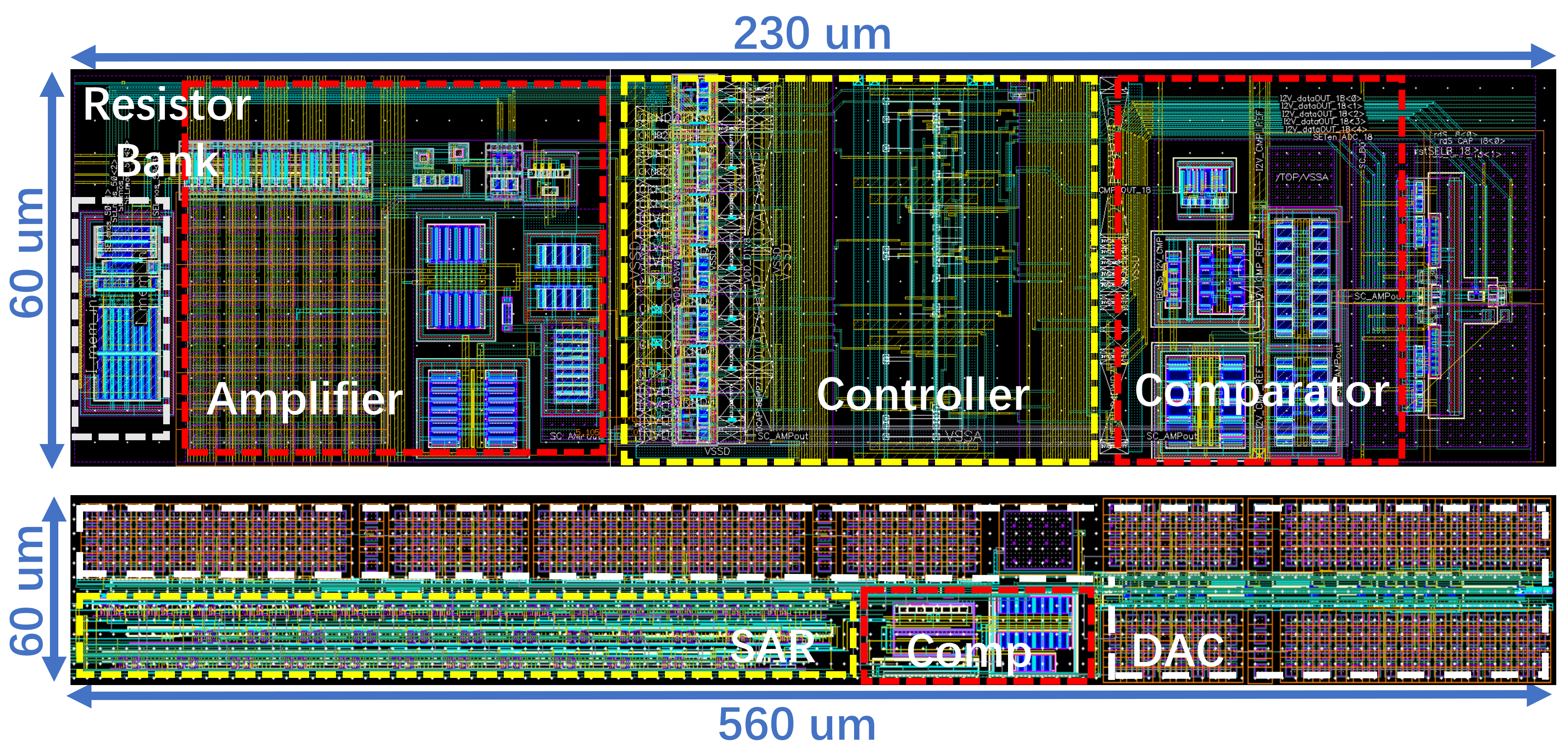}
  \caption{\footnotesize Layout. Upper: I-to-V circuit; Lower: SAR ADC}
  \label{Fig:9}
\end{figure}

This section shows the final layout (Fig. \ref{Fig:9}) of the read-out circuit done in Cadence with TSMC 180nm technology. 
The height is unified so that the two blocks can be combined and put in the same row/column, paving the path for integrating multiple read-out circuits onto a memristor crossbar.

\begin{table}[!t]
\centering
\caption{Specifications of proposed design  }
 \begin{center}
 \begin{tabular}{lcc} 
     \toprule
     Specifications & Unit & Value  \\  
     \midrule
     Voltage supply & (V) & Resistor bank: 5/ Others: 1.8 \\ 
     \midrule
    Circuit size & ($\mu$m) & 60$\times$790 \\
     \midrule
     INL & (LSB) &   +0.81/-0.75\\
      \midrule
      DNL & (LSB) &  +0.94/-0.44\\
       \midrule
        ADC sampling frequency & (kHz) &250\\ 
     \midrule
        Read-out frequency  & (kHz) &50-200\\ 
     \midrule
     Average power consumption  & ($\mu$W)& 536.07 \\

    \bottomrule
\end{tabular}	
\label{table:mm}
\end{center}
\end{table}

\section*{Conclusion}
\label{3}

Conventional circuits encountered the issue of reading-out and digitising the wide-ranging memristor conductance/current with high accuracy and linearity. This issue is solved by a novel read-out circuit proposed in this paper. Input currents (20nA-2mA) can be dynamically divided into 5 measuring scales, converted and amplified to voltages (160mV-1.2V), and digitised. This self adjusting read-out circuit significantly reduces errors and non-linearity especially when the measured current is too small or too large. The read-out accuracy and linearity can be further improved with more resistors in the resistor bank with the price of longer operating time and more energy. In summary, this proposed design is promising in memristor based memory and computing which requires large output range and high precision.


\vspace{12pt}
\color{red}


\begin{thebibliography}{00}

\bibitem{chua}
L.O. Chua. Memristor-the missing circuit element. Circuit Theory, IEEE Transactions on,
18(5):507-519, 1971. ISSN 0018-9324. doi: 10.1109/TCT.1971.1083337.

\bibitem{HP}
R. S. Williams, "How We Found The Missing Memristor," in IEEE Spectrum, vol. 45, no. 12, pp. 28-35, Dec. 2008, doi: 10.1109/MSPEC.2008.4687366.

\bibitem{atomic_switch}
M. Aono and T. Hasegawa, "The Atomic Switch," in Proceedings of the IEEE, vol. 98, no. 12, pp. 2228-2236, Dec. 2010, doi: 10.1109/JPROC.2010.2061830.

\bibitem{RRAM}
C. Xu, X. Dong, N. P. Jouppi and Y. Xie, "Design implications of memristor-based RRAM cross-point structures," 2011 Design, Automation and Test in Europe, 2011, pp. 1-6, doi: 10.1109/DATE.2011.5763125.

\bibitem{multi_bit}
H. Kim, M. P. Sah, C. Yang and L. O. Chua, "Memristor-based multilevel memory," 2010 12th International Workshop on Cellular Nanoscale Networks and their Applications (CNNA 2010), 2010, pp. 1-6, doi: 10.1109/CNNA.2010.5430320.

\bibitem{face}
P. Yao, H. Wu, B. Gao, S. B. Eryilmaz, X. Huang, W. Zhang, Q. Zhang, N. Deng, L. Shi,H.-S. P. Wong et al., “Face classification using electronic synapses,” Nature communications, vol. 8, p. 15199, 2017, doi: 10.1038/ncomms15199

\bibitem{fully2}
Z. Wang et al., “Fully memristive neural networks for pattern classification with unsupervised learning,” Nature Electronics, vol. 1, pp. 137–145, 2018.
  
\bibitem{image}
C.  Li,  M.  Hu,  Y.  Li,  et al., “Analogue signal and image processing with large memristor crossbars,” Nature Electronics, vol. 1, 01 2018.

\bibitem{efficient}
C. Li, D. Belkin, Y. Li, P. Yan, M. Hu, et al.,“Efficient and self-adaptive in-situ learning in multilayer memristor neural networks,”Nature Communications, vol. 9, 2018.

\bibitem{dot-product}
M. Hu et al., “Memristor-based analog computation and neural network classification with a dot product engine,” Advanced Materials, vol. 30, 2018.

\bibitem{fully}
F.  Cai,  J.  Correll,  S.  H.  Lee,  Y.  Lim,  V.  Bothra,  Z.  Zhang,  M.  Flynn,  and  W.  Lu,  “A fully integrated reprogrammable memristor–cmos system for efficient multiply–accumulate operations,” Nature Electronics, vol. 2, p. 1, 07 2019.

\bibitem{hardware}
 P.  Yao,  H.  Wu,  B.  Gao,  J.  Tang,  Q.  Zhang,  W.  Zhang,  J.  Yang,  and  H.  Qian,  “Fully hardware-implemented  memristor  convolutional  neural  network,” Nature,  vol.  577,  pp.641–646, 2020.
 
\bibitem{ADC}
B. P. Ginsburg and A. P. Chandrakasan, "An energy-efficient charge recycling approach for a SAR converter with capacitive DAC," 2005 IEEE International Symposium on Circuits and Systems, 2005, pp. 184-187 Vol. 1, doi: 10.1109/ISCAS.2005.1464555.

\end{thebibliography}
\end{document}